\documentclass{article}
\usepackage{arkiv}
\usepackage{cite}
\usepackage{amsmath,amssymb,amsfonts}
\usepackage{algorithmic}
\usepackage{graphicx}
\usepackage{textcomp}

\usepackage{listings}
\usepackage[frozencache=true,cachedir=.,newfloat=false]{minted}
\newcommand\minty[1]{\texttt{#1}}
\newcommand\citep[1]{\cite{#1}}
\usepackage{booktabs}
\usepackage{hyperref}

\title{Procedural Generation of STEM Quizzes}
\author{Carlos Andujar \\
ViRVIG, Computer Science Dept. \\
Universitat Politècnica de Catalunya \\
Barcelona, 08034 Spain \\
\texttt{andujar@cs.upc.edu}
}

\begin{document}
\maketitle

\begin{abstract}
Electronic quizzes are used extensively for summative and formative assessment. Current Learning Management Systems (LMS) allow instructors to create quizzes through a Graphical User Interface. Despite having a smooth learning curve, question generation/editing process with such interfaces is often slow and the creation of question variants is mostly limited to random parameters. In this paper we argue that procedural question generation greatly facilitates the task of creating varied, formative, up-to-date, adaptive question banks for STEM quizzes. We present and evaluate a proof-of-concept Python API for script-based question generation, and propose different question design patterns that greatly facilitate question authoring. The API supports questions including mathematical formulas, dynamically generated images and videos, as well as interactive content such as 3D model viewers. Output questions can be imported in major LMS.  For basic usage, the required programming skills are minimal. More advanced uses do require some programming knowledge, but at a level that is common in STEM instructors. A side advantage of our system is that the question bank is actually embedded in Python code, making collaboration, version control, and maintenance tasks very easy. We demonstrate the benefits of script-based generation over traditional GUI-based approaches, in terms of question richness, authoring speed and content re-usability.
\end{abstract}


\section{Introduction}
\label{sec:introduction}
Quizzes are valuable tools for education and assessment. Although some quizzes are still delivered on paper, we focus on electronic quizzes, since they are environmentally friendly, support online delivery, and fully benefit from the display, editing, tracking, evaluation, reporting and analysis features of Learning Management Systems \cite{gamage2019}.

Quizzes consist of a collection of questions of different types. Most question types (e.g. multiple choice, true-false, short answer) allow the LMS to automatically evaluate and grade the students' responses. 
Computer-scored quizzes provide scalability (since grading is automatic, they can be applied to arbitrarily-sized groups and even Massive Open Online Courses), objectivity in scoring (in contrast to essays, which are subject to different degrees of scoring errors and scorer bias) and immediate feedback (attempts, whether correct, marks, right answers). Due to these features, electronic quizzes are an effective tool for continuously monitoring student performance. Such quizzes can be used both for high-stakes activities (like mid-terms and finals), and frequent, low-stakes tests for self-assessment.
 
\begin{figure}
	\centering
		\includegraphics[width=0.9\columnwidth]{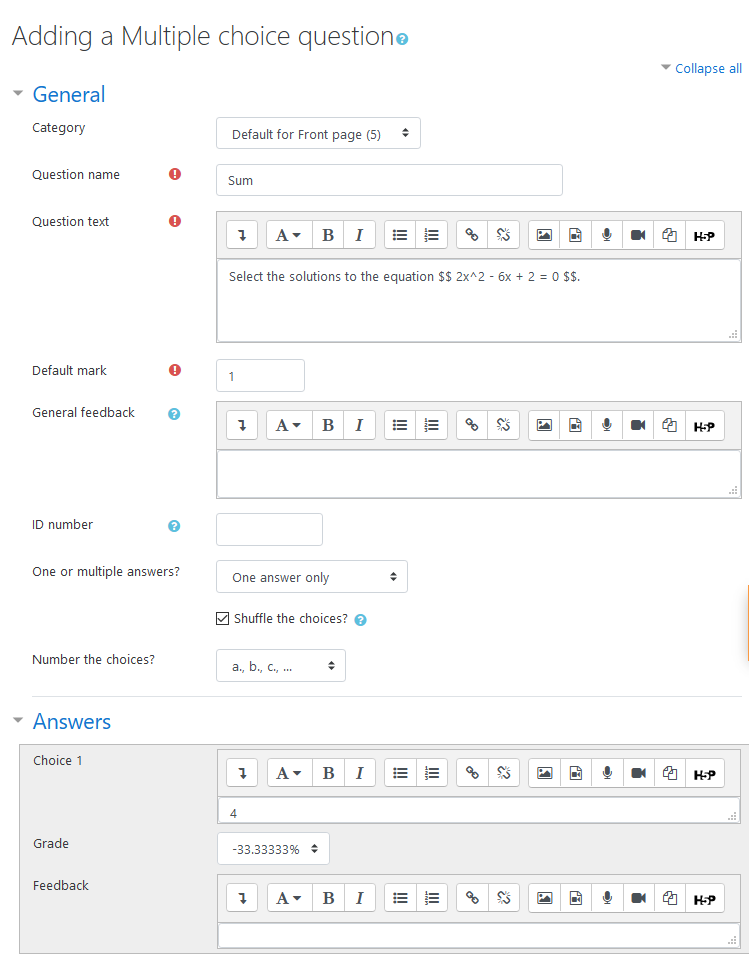}
	\caption{Part of Moodle's form for creating a multiple-choice question.}
	\label{fig:moodleform}
\end{figure}

The traditional approach is to create questions using a Graphical User Interface  (Figure~\ref{fig:moodleform}). Questions can be grouped into hierarchical categories and kept into a question bank so that they can be re-used for different quizzes. Categories often refer to course units, difficulty levels or academic semesters. The ability to create a large question bank carries substantial educational advantages: better coverage of course topics, broader question goals (recall, comprehension, application\dots), more distinct practice tests, and more opportunities to prevent cheating. 
Two major strategies to prevent cheating on online quizzes are timed quizzes (to keep students from looking up answers in textbooks) and randomization (to prevent question sharing by randomizing question order, choice order, and letting quizzes include random subsets from the question bank). The usage of random questions is promoted in current LMS to minimize the potential for cheating \citep{manoharan2016} and, most importantly, to increase the opportunity for students to learn from the feedback provided by repeated quiz attempts \citep{calm2017,forster2018}. The idea is that instructors create slight variants of each question, and define quizzes by including random subsets, so that each student gets one of the variants picked at random. Categories can be used to group questions covering the same topic with similar difficulty. 

Current LMS offer limited capabilities for creating large question banks in a time-effective manner. Most LMS provide a question-creation GUI consisting of a variety of form fields for defining the category, question name, question text, answers, default mark, feedback and tags (Figure~\ref{fig:moodleform}). For example, Moodle's multiple-choice question form includes (for a 4-choice question) more than 25 fields.  
Despite these forms provide substantial flexibility, creating a large number of questions, or variants of a given question, takes a considerable amount of time.  Furthermore, embedding images into questions (e.g. electrical circuits, function plots) lacks scalability because the images must be created with external tools and then imported into the LMS. 

In this paper we argue that a one-for-all form-based GUI might not be the optimal interface for question generation, and that STEM instructors with programming skills can largely benefit from procedural question generation. We propose and evaluate a script-based method for fast creation of common quiz question types. Our approach is based on a proof-of-concept Python library to create single questions (multiple-choice, numerical, short-text\dots) and random variants from user-provided lists. Besides the API, we discuss different question design strategies to create question variants at multiple levels of similarity.
 
Listing~\ref{list:first_example} provides a first-contact with the API, consisting of a script example, and some random questions generated by the script. Since our approach requires some programming skills to generate questions, we restrict ourselves to STEM (science, technology, engineering, and mathematics) courses, since STEM instructors often have programming skills or experience with scientific software such as MatLab.

\begin{listing*}[ht]
\begin{minted}[fontsize=\footnotesize]{python}
# Create a few questions about derivatives
from quizgen import *

Q = Quiz('listing1.xml')
x = symbols('x')
functions = [ cos(x**2), 2*x*sin(x), sin(x)*cos(x), 2*sin(cos(x)), sin(2*x), tan(2*x) ]
pairs = [ (f, diff(f))  for f in functions] # (f, f') pairs
Q.addMultipleChoiceFromPairs("Derivatives", "Select the derivative of %s:", pairs)
Q.preview()
Q.close()
\end{minted}
\centering
		\includegraphics[height=0.20\textwidth]{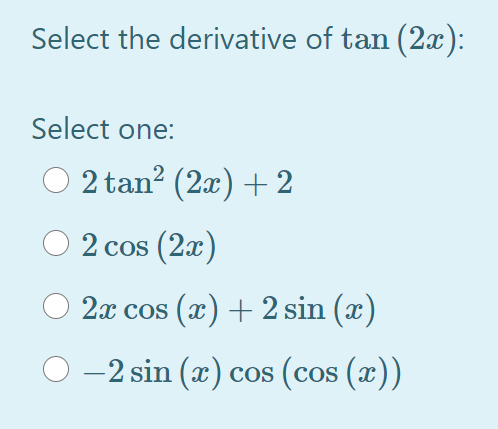}
		\includegraphics[height=0.20\textwidth]{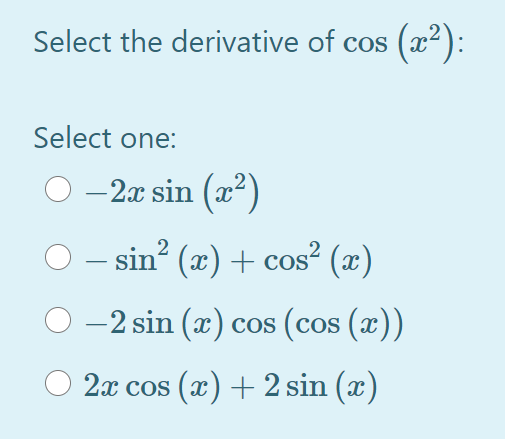}
		\includegraphics[height=0.20\textwidth]{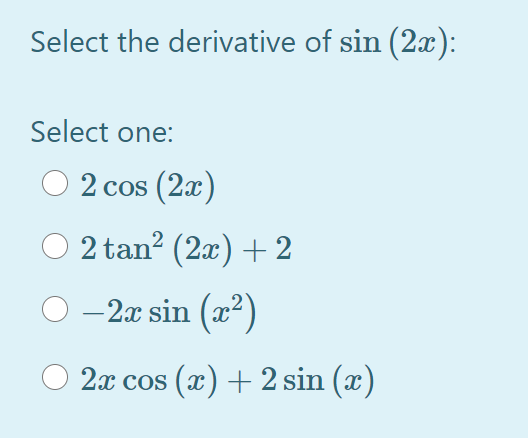}
		\includegraphics[height=0.20\textwidth]{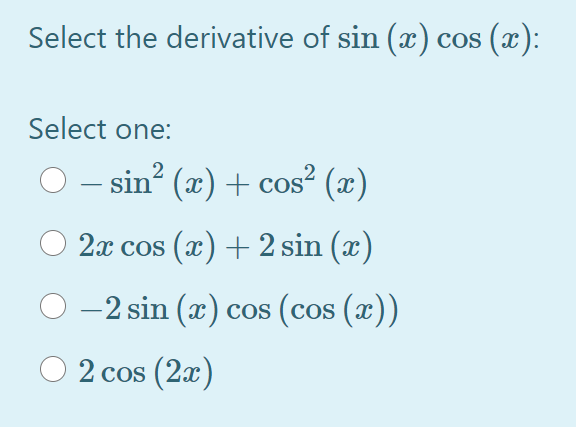}
\caption{Sample Python script using the proposed API to create some calculus questions. This example uses SymPy library~\cite{meurer2017} to define and differentiate some trigonometric functions. }  
\label{list:first_example}
\end{listing*}

We analyzed and compared our approach with Moodle, a popular and open-source LMS. We compared question editing times for both systems through a user study. 
Our experiments show that the script-based approach clearly outperforms Moodle's GUI in terms of question creation speed and easiness to create rich, varied,   compelling questions with dynamically-generated values, formulas, images and videos.

The rest of the paper is organized as follows. Section~II reviews related work including GUI-based question creation and randomness generation in current LMS.
We describe the proposed Python API in Section~III, and discuss specific question design guidelines in Section~IV. We provide further examples in Section~V. We evaluate the script-based approach in Section~VI, by comparing instructor performance in authoring and maintenance tasks with GUI-based and script-based approaches. Finally, Section~VII concludes the paper and outlines future research avenues. 
 
\section{Previous work}
 
\subsection{Question creation in LMS}
In current LMS, questions can be added via online forms or by importing questions in multiple formats. Examples of text-based question formats include Aiken (multiple-choice questions), GIFT (multiple-choice, true-false, short answer, matching missing word and numerical questions) and Cloze (multiple-choice, multiple-answer questions with embedded answers). 

Although there are many free and commercial LMS available, we will focus our examples on Moodle, a free and open-source LMS with +200 million users from +240 countries at the time of this writing. Other popular LMS are Blackboard Learn \citep{patterson2013} and WebCT \citep{ngai2007}, which also provide file formats for question import/export operations. 
Besides general LMS, some tools target specifically question generation. Most of these tools  (e.g. ExamView, 
Pearson's TestGen) 
allow users to export questions so that they can be re-used in major LMS.

All these tools provide form-based interfaces for question creation. These interfaces put emphasis on ease of use and thus can be used by instructors with a variety of profiles. 
Figure~\ref{fig:moodleform} shows Moodle's form for creating a multiple-choice question. Forms help users understand the meaning of the different question fields and settings, which lets novice users create questions in a few minutes. Unfortunately, the content for the different form fields (e.g. question text and choices) must be typed separately for each question, with no options for re-use besides duplication of the whole question. Consider for example the multiple-choice questions shown in Listing~\ref{list:first_example}. The questions share a clear pattern (both question text and answers) but form-based interfaces in existing tools do not facilitate content re-use between questions. For the questions in Listing~\ref{list:first_example}, users would need to duplicate questions manually and then edit both the varying part of the question text and the grade associated to the choices (which is prone to errors).

\subsection{Automatic question/answer generation}

Current LMS offer limited features to create new questions automatically from existing questions. For example, Moodle supports the creation of random matching questions by picking randomly a few short-answer questions from the question bank. 

A more powerful question type is calculated question, where the question text can include variables e.g.  $\{x\}$, $\{y\}$ that will be substituted by a numerical value from their corresponding dataset. The answers can include mathematical expressions using these variables. For example, a question text might read "Compute \{a\}+\{b\}" and the correct answer can be written as the formula "\{a\}+\{b\}". The system allows users to associate variables with specific datasets; each dataset is described by a range of values, decimal places, and distribution (e.g. uniform). GUIs for calculated questions are often more cumbersome than fixed questions, but once created, the same question with different numerical parameters can be reused in multiple quizzes. Although Moodle's formulas can include both arithmetic operators and mathematical functions, no symbolic computation is supported and thus the questions in Listing~\ref{list:first_example} cannot be represented as a calculated question. 

Some approaches focus on extending this concept of calculated questions. WIRIS \citep{ferrer2002} provides a server-side SDK for calculus and algebra that can be integrated with LMS through WIRIS quizzes \citep{mora2011}.
WIRIS has been applied with great success in math-related courses, see e.g. \citep{rodriguez2012,calm2017}. As in our approach, advanced WIRIS usage requires programming skills. A major advantage of WIRIS quizzes is that students answers may consist in formulas ---entered using an integrated editor--- and that answers are checked for multiple types of mathematical equivalence when grading the quizzes. This, however, comes at the cost of having to commit a great deal of computing power for the server to support medium to large groups of students being tested simultaneously. Open answers (e.g. "type a prime number larger than 100") are also supported through WIRIS's own programming language. A similar approach is adopted by Maple TA, which evaluates student answers to check for mathematical equivalence with the correct answer. Maple TA questions can be generated through a GUI and also through LaTeX \citep{heck2004,jones2008,pereira2010}. 
We further extend these ideas by providing an API to create arbitrary quiz questions using a Python script. This facilitates random question creation for arbitrary fields beyond mathematics.   

CodeRunner \citep{lobb2016} is a Moodle plugin that allows instructors to run a program in order to grade answers. This type of question is useful in programming courses where students are asked to write code according to some specification, which is then graded by running a test set. Besides programming quizzes, CodeRunner can be used also for questions having many different correct answers and whose correctness can be assessed by a program.  CodeRunner supports major programming languages including Python, C++, Java, JavaScript and Matlab. CodeRunner has been extended e.g. to handle OpenGL assignments, including interactive 3D renderings \citep{wunsche2019}. Although our approach also benefits from programming, both systems pursue different goals and with different strategies. Our system runs instructor-provided scripts on the instructor's computer to create questions, rather than running student code on sandboxes of a LMS server each time a student submits a response. 

There are many other web-based coding tools and online judges, see e.g.  \citep{Kurnia2001,Petit2012,Fox2015,Povzenel2015,Rajala2016,benotti2018,Andujar2018} 
but they address exclusively programming exercises and do not integrate seamlessly with current LMS. 
Some approaches (e.g. UCLA's Question Bank Quick Builder) allow users to enter questions into spreadsheets, which are later exported to an LMS format, or generate them automatically from an ontology~\cite{wang2019}. 
These approaches greatly speed-up question typing, but are mostly limited to text-based questions.

\section{API description}

The API provides methods to create common LMS question types: multiple-choice, numerical, short-answer and matching questions. Listing~\ref{list:high-level} shows the most relevant methods; see the accompanying source code for further details. We first describe the methods that create single questions, and then the methods that facilitate the automatic generation of random questions from user-provided lists. 

\begin{listing*}[ht]
\begin{minted}[fontsize=\footnotesize]{python}
class Quiz:
  # add single questions
  def addShortAnswer(name, question, answer)
  def addNumerical(name, question, answer, tolerance=0.01)
  def addMultipleChoice(name, question, choiceList)
  def addMatching(name, question, pairList)
  # add random questions from lists
  def addMultipleChoiceFromLists(title, question, correctAnswers, distractors, numQuestions=-1)
  def addMultipleChoiceFromPairs(title, question, pairs, moreDistractors=[], numQuestions=-1)
  def addCompleteCode(title, question, sourceCode, tokens, distractors = [], numQuestions=-1)
  def preview()
  def close()
\end{minted}
\caption{Proof-of-concept API for creating quizzes (some methods have been omitted).
} \label{list:high-level}
\end{listing*}


The methods \minty{addShortAnswer}, \minty{addNumerical}, \minty{addMultipleChoice} and \minty{addMatching} allow the creation of single questions, and constitute the building blocks for more advanced methods generating multiple questions. Listing~\ref{list:low-example} shows basic usage examples. Since our quizzes are meant to be presented with web-based LMS, we support HTML in questions and answers. We also support LaTeX code (e.g. for equations). 


        
\begin{listing*}[ht]
\begin{minted}[fontsize=\footnotesize]{python}
# Create some simple questions
from quizgen import *

Q = Quiz('listing3.xml')
# Add a pair on questions on quadractic equations
a, b, x = randint(2,6), randint(1,9), randint(3,8)
c = -a*x**2 - b*x
solutions = [x, float((-b-sqrt(b*b-4*a*c))/(2*a))]
distractors = list(set([x-1, x+1, x-2, x+2]) - set(solutions))
Q.addNumerical("", f"Solve \( {a}x^2+{b}x{c}=0 \)", solutions)
Q.addMultipleChoice("", f"Select a solution for \({a}x^2+{b}x{c}=0\)", [x]+distractors)
# Add a question on radiometry units
pairs = [("Flux", "W"), ("Intensity", "W/sr"), ("Irradiance", "W/m^2"), ("Radiance", "W/(sr*m^2)")]
Q.addMatching("", "Match magnitudes with units:", pairs)
Q.close()
\end{minted}
\centering
\includegraphics[width=0.9\textwidth]{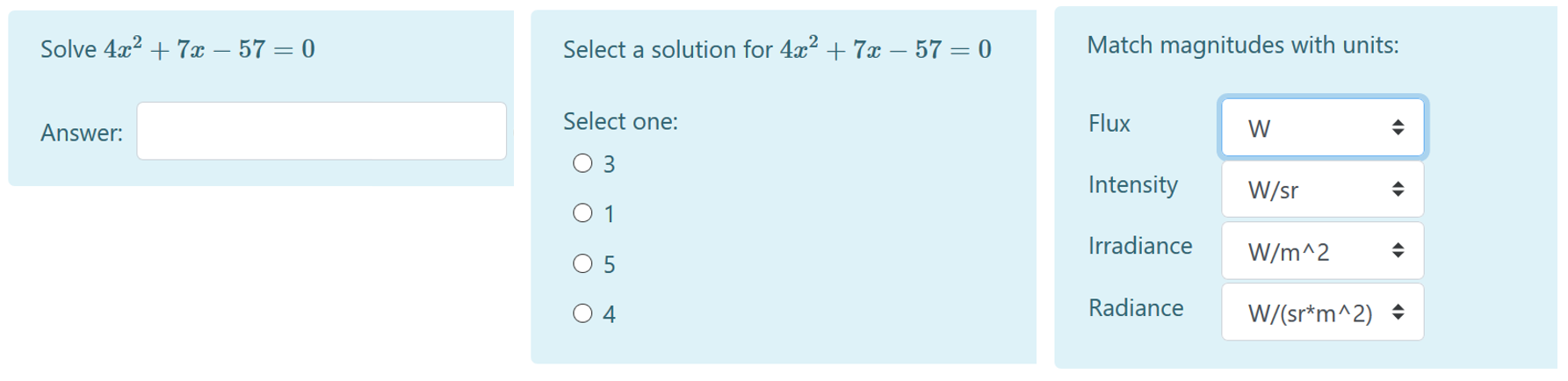}
\caption{Simple example for creating single questions.} 
\label{list:low-example}
\end{listing*}

The API also provides methods to facilitate the automatic generation of random questions from user-provided lists. 
The first option (\minty{addMultipleChoiceFromLists}) requires instructors to provide two lists: one with correct answers and another with wrong answers (distractors). Listing~\ref{list:multi1} shows one example. 
The API picks a random choice from the correct answer list, and the rest of options (usually three more) are picked from the distractor list. The method checks that all choices are distinct, otherwise a warning is issued and no question is added. 
The second method (\minty{addMultipleChoiceFromPairs}) requires instructors to provide a list with (key, answer) pairs. Listing~\ref{list:multi2} shows one example where \textit{keys} are equation parts and \textit{answers} are their meaning. 

\begin{listing*}[ht]
\begin{minted}[fontsize=\footnotesize]{python}
# Sample script using addMultipleChoiceFromLists
from quizgen import *
Q = Quiz('listing4.xml')
# Add some questions
onlyVS = ["Write gl_Position.", "Write to an out variable with texture coordinates.",
    "Animate the geometry of the 3D model.", "Compute per-vertex lighting."]
onlyFS = ["Call dFdx, dFdy functions.", "Execute discard.", "Write fragColor.", "Read gl_FragCoord.",
    "Write gl_FragDepth.", "Apply bump mapping.", "Apply normal mapping."]
both = ["Compute the light vector.", "Compute lighting."]
none = ["Write to gl_FragCoord.", "Create new primitives.", "Create new fragments."]
question = "Select the task that makes sense in a GLSL "
Q.addMultipleChoiceFromLists("", question + "<b>Vertex Shader</b>:", onlyVS + both, onlyFS + none)
Q.addMultipleChoiceFromLists("", question + "<b>Fragment Shader</b>:", onlyFS + both, onlyVS + none)
\end{minted}
\centering
		\includegraphics[width=0.9\textwidth]{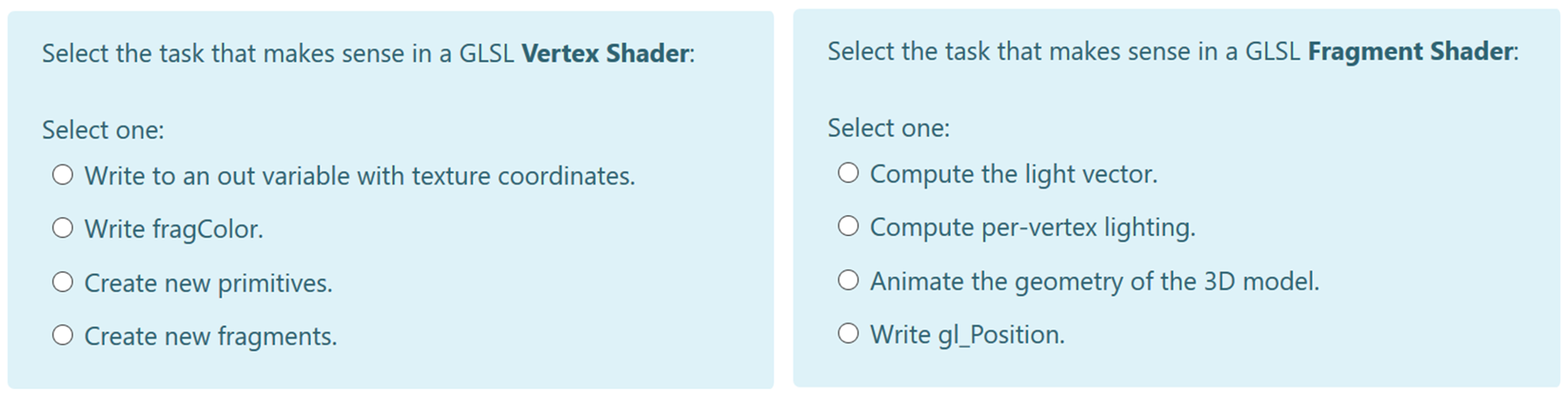}
\caption{Adding multiple-choice questions from user-provided lists.
} \label{list:multi1}
\end{listing*}

\begin{listing*}[ht]
\begin{minted}[fontsize=\footnotesize]{python}
# Sample script using addMultipleChoiceFromPairs
from quizgen import *
Q = Quiz('listing5.xml')
# Add questions on Kajiya's rendering equation
Lo = "L_o(x, \omega_o, \lambda ,t)"
Le = "L_e(x, \omega_o, \lambda ,t)"
Li = "L_i(x, \omega_i, \lambda ,t)"
fr = "f_r(x, \omega_i, \omega _o, \lambda,t)"
dot = "(\omega_i \cdot n)"
equ = f"$${Lo} = {Le}\ + \int_\Omega {fr}{Li}{dot}d\omega_i$$"
question = f"Kajiya's rendering equation can be written in the form {equ}. <p> What is \(%s\)?"
keyAnswerPairs = [
    (Lo, "Exiting radiance."),  (Le, "Emitted radiance."),
    (Li, "Incident radiance."), (fr, "Material's BRDF."),
    (dot, "Cosine of incident angle."),
    ("\lambda", "Radiant energy wavelength."),
    ("\Omega", "Unit hemisphere.")]
distractors = ["Irradiance.", "Illuminance.", "Intensity.","Incident direction."]
Q.addMultipleChoiceFromPairs("", question, keyAnswerPairs, distractors)
\end{minted}
\centering
\includegraphics[width=0.8\textwidth]{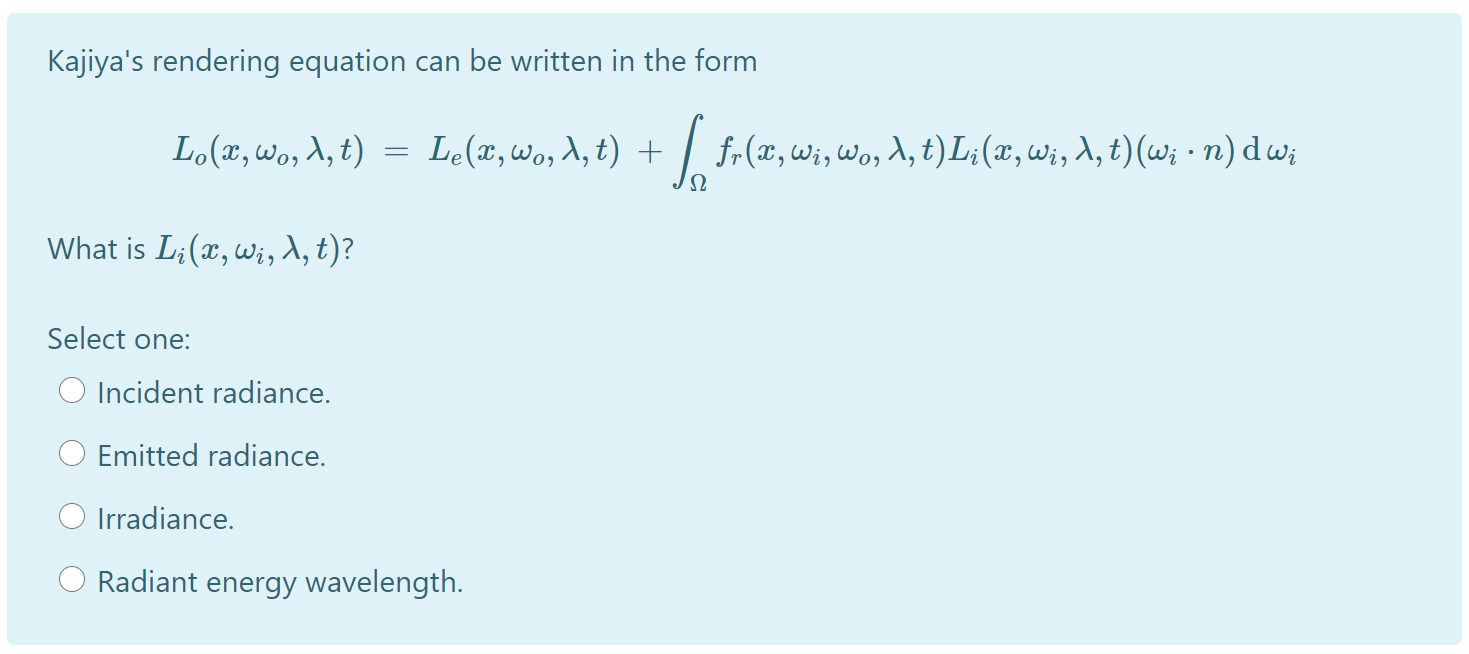} 
\caption{Adding multiple-choice questions from (key, answer) pairs. In this example, \textit{key} is a part of an equation, and \textit{answer} is its interpretation.} 
\label{list:multi2}
\end{listing*}

\section{Question design guidelines}

This section provides guidelines and question design strategies to fully benefit from the proposed API. In the following discussion, we distinguish between the information needed to define a set of questions (that we will refer to as \textit{question corpus}) and the specific question formats we might choose to ask about the corpus (\textit{question structure}). For example, a collection of (city, country) pairs might define a corpus, from which we could ask  true-false questions ("Is $X$ city located in $Y$? ), short-answer questions ("What country is city $X$ located in?"), or matching questions ("Match each city with the country it is located in").
 
\subsection{Multiple-choice questions from answer and distractor lists}
 
The method \minty{addMultipleChoiceFromLists} allows instructors to generate questions that require students to check if items in the alternatives fulfill some fixed Boolean property. We have shown one example where the items were programming tasks, and the property was whether the task makes sense in a specific shader or not (Listing~\ref{list:multi1}). These questions consist of a fixed stem (stating the property) and four items; one of them fulfilling the property, and the others not.  
 
A few examples of items and properties that can be used to define this type of questions:
\begin{itemize} 
\item For function items: is invertible? is differentiable? is continuous? is monotonic? has critical points?
\item For matrix items: is singular? is positive-definite? is orthogonal?  
\item For systems of equations: is consistent? is overdetermined? is underdetermined?
\end{itemize}
 
More formally, the question corpus for these questions is determined by a set $S=\{a_i\}$ of items (e.g. functions, matrices, tasks) together with some Boolean property $P(a_i)$. In the proposed API, the instructor has to provide two sets: $S_c = \{ a\in S\ | P(a) \}$ with correct items (fulfilling the property), and distractors $S_d \subset S - S_c$. For small item collections, the sets can be provided manually as part of the Python script, as in Listing~\ref{list:multi1}.  For larger collections, a better strategy is to use an algorithm to generate a random collection of items (e.g. polynomials, matrices\dots) together with a function $f_P(a_i)$ for checking the property. An important difference with respect to related approaches (e.g. WIRIS quizzes) is that function $f_P(a_i)$ can benefit from the myriad of scientific Python packages, and that it is evaluated at quiz generation time and thus it involves no runtime overhead every time a student submits an answer.
 
We now discuss the number of distinct questions that can be generated from $S_c$ and $S_d$. Let $c$ be $|S_c|$ and $d$ be $|S_d|$. We say two multiple-choice questions are \textit{unique} if they have different correct answers; distractors might overlap. Two multiple-choice questions are \textit{distinct} if at least one choice is different. Choice order is not considered in this paper, since the quiz will be exported to a LMS (e.g. Moodle) that will take care of proper randomization of choices within questions, and questions within quizzes. Since we have $c$ correct answers and $d$ distractors, we can generate up to $c$ unique questions, and up to $c\binom{d}{3}$ distinct questions. The last parameter of \minty{addMultipleChoiceFromLists} specifies how many questions to add; by default, we add $c$ unique questions.
 
When using this method, instructors must recheck that the both sets $S_c$ and $S_d$  are well defined. If by mistake some element $a\in S_c$ does not fullfill $P(a)$, some questions might have no solution, which might confuse students and be a major problem in timed quizzes.
Conversely, if some distractor $a\in S_d$ does fullfill $P(a)$, some questions might include multiple choices that are theoretically correct, but one of them will not be recognized as such and thus receive a penalty.
For high-stake tests, instructors should also check that the student effort for testing the property (both directly or through elimination) is similar for all items in the collection. Finally, as suggested in the literature~\cite{burton1990}, distractors should be plausible, e.g. showing common student misconceptions. Easy-to-check items should not be included as distractors. For example, if students must identify which option is a prime number, distractors should exclude multiples of 2 and 5, as in Listing~\ref{list:primes}. 

\begin{listing*}[ht]
\begin{minted}[fontsize=\footnotesize]{python}
# Sample script using answer + distractor lists
from quizgen import *
Q = Quiz('listing6.xml')
# Add question asking to identify a 3-digit prime number
M=999
sieve.extend(M)
prime_numbers = list(sieve._list) # prime numbers up to 999
multiples_of_2 = set(range(0, M, 2))
multiples_of_5 = set(range(0, M, 5))
distractors =  list(set(range(100,M)) - set(prime_numbers) - multiples_of_2 - multiples_of_5)
Q.addMultipleChoiceFromLists("", "Select the <b> prime </b> number:", prime_numbers[25:], distractors, 5)
\end{minted}
\centering
\includegraphics[width=0.8\textwidth]{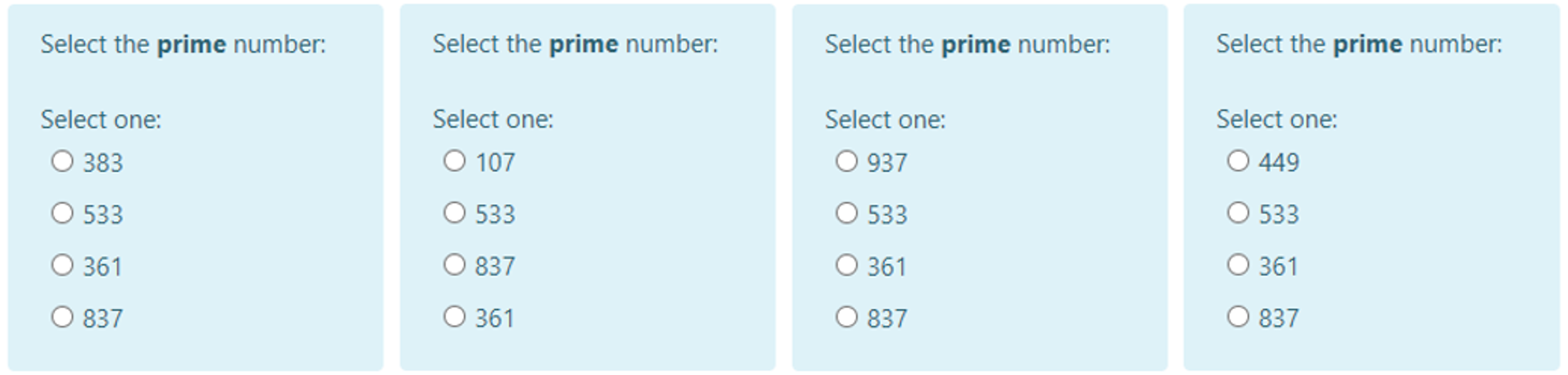} 
\caption{Multiple-choice questions from answers/distractor lists. The script generates prime numbers and plausible distractors.} 
\label{list:primes}
\end{listing*}
 
\subsection{Multiple-choice questions from key-answer pairs}

The method \minty{addMultipleChoiceFromPairs} is suitable for multiple-choice questions whose stem has a fixed part (referred to as \textit{question pattern}) and a varying part (\textit{question key}) that conditions the correct \textit{answer}. The user provides a set of (key, answer) pairs.
For each random question, our system will pick randomly one (key, answer) pair, and three answers from other pairs. An additional distractor list can be provided, so that distractors are taken both from other answers and the distractor list. 
The question is formed by inserting the key text in the question pattern (we use \textit{\%s} in the question pattern to indicate where the key should appear), and presenting the four choices (one of them being correct), see Listing~\ref{list:first_example}. 
 
Examples of pairs that can be used to create these types of questions include (city, country) pairs, (function, derivative) pairs, (matrix, eigenvalues) pairs and (equation, solution) pairs. 
Instructors should pay attention to define (key, answer) pairs such that, for any given key, there is only one valid answer in the rest of pairs and the  distractors. For example, the set $\{$ ('Lyon','France'), ('Marseille', 'France'), \dots$\}$ defines a non-injective function. Our current implementation accepts the above input, but checks, for each question being generated, that none of the distractors matches the chosen answer. However, this check is currently based on a string-based comparison, and thus a set such as $\{$ ('Beijing', 'China'), ('Shangai', "People's Republic of China"), \dots$\}$  will result in confusing alternatives.

Pairs can be created manually as in Listing~\ref{list:multi2}, 
or programmatically as in Listing~\ref{list:first_example}. Injectivity can be achieved in multiple ways. For shorter (key, answer) sets, the injectivity condition is easy to verify manually. For larger sets, a simple option is to generate potentially non-injective pairs, and then use domain knowledge to remove pairs with duplicated answers. This involves having a function that checks for answer equivalence. For example, we can create random polynomials and use a symbolic package (e.g. SymPy~\cite{meurer2017}) to compute their derivatives. Then, we can use the same package to check for mathematical equivalence and filter duplicate answers. This task is very domain-specific and thus is part of the data preparation before actually calling our API method.
 
Concerning the number of questions: let $c$ be the number of (key, answer) pairs, and $d$ the total number of distractors for a given key ($d=c-1+a$, where $a$ is the number of additional distractors). We can generate up to $c$ unique questions, and up to $c\binom{d}{3}$ distinct questions. Again, our default is to generate $c$ unique questions.
 
Some server-side systems (e.g. WIRIS quizzes) have built-in functionalities that check the student's answer in a mathematical fashion. This allows students to write equations in the answer field, and the system will check the mathematical equivalence with the instructor-provided solution (in our prototype short-answer questions are limited to answers that can be checked with a simple string comparison). On the other hand, our system is more general in the sense that it can be applied to arbitrary objects within the Python script. Again, since our equivalence checks are done during quiz generation time, the questions we generate have no performance overhead during exams.

\subsection{Numerical questions with random content}
 
Although we encourage extensive use of this question generation strategy, our API just provides a method for adding single numerical questions. However, we have found this to be powerful enough in the context of a Python script. 
The method \minty{addNumerical} requires a question and a list with multiple numerical answers. For example, we might ask for a solution of a quadratic equation, and provide two possible solutions, as in Listing~\ref{list:low-example}. This obviously works for a finite (and relatively small) number of solutions. If that is not the case, 
one strategy is to modify the question to ask for solutions in a certain interval. For example, Listing~\ref{list:primes} can be trivially edited to add  numerical questions by adding:

\begin{minted}[fontsize=\footnotesize]{python}
Q.addNumerical("", "Enter a 3-digit prime number:", 
prime_numbers[25:])
\end{minted}
{
}

The recommended approach for generating each of these questions is to first generate random values to instantiate a random object (e.g. polynomial, matrix), then create the question text using Python f-strings, and finally coding the rule that computes the answers given an object instance. Python f-strings (e.g. f"Compute \{a\}+\{b\}") are available since Python 3.6. They provide a very convenient way for writing questions where some parts should be replaced by arbitrary expressions that are evaluated at runtime.
 
Instructors need to recheck that the question text is written correctly (e.g. no missing curly braces in the f-strings). Fortunately, the automatic preview feature of our prototype (discussed below) greatly simplifies this review task.

 \subsection{Fill-in-the-blanks questions from a text and a token list}
 
This method is provided by \minty{addCompleteCode}. 
The instructor must provide some text (e.g. source code) along with a list of tokens (arbitrary strings appearing in the text). Listing~\ref{list:complete} shows one example.
The method picks randomly a token from the list, and replaces all its appearances in the text by "$\rule{1cm}{0.15mm}$". The question is formed by adding the instructor-provided text to a question pattern indicating that one must select the more suitable choice to fill the blank indicated by "$\rule{1cm}{0.15mm}$". The alternatives include the correct choice (token being replaced), and distractors taken from other tokens. As in some previous methods, the instructor might provide a list of additional distractors.

\begin{listing*}[ht]
\begin{minted}[fontsize=\footnotesize]{python}
# Sample script for fill-in-the-blanks questions
from quizgen import *
Q = Quiz('listing7.xml')
code = """
    void main()
    {
        vec3 P = (modelViewMatrix * vec4(vertex, 1.0)).xyz;
        vec3 N = normalize(normalMatrix * normal);
        vec3 V = normalize(-P);
        vec3 L = normalize(lightPosition.xyz - P);
        frontColor = PhongLight(N , V , L);
        gl_Position = modelViewProjectionMatrix * vec4(vertex, 1.0);
    }
"""
tokens = ['modelViewMatrix', 'modelViewProjectionMatrix', "normalMatrix"]
distractors = ['viewMatrix', 'viewProjectionMatrix', 'modelViewMatrixInverse']
Q.addCompleteCode("", "Complete this vertex shader: <p> <pre>%s</pre>", code, tokens, distractors)
\end{minted}
\centering
\includegraphics[width=0.8\textwidth]{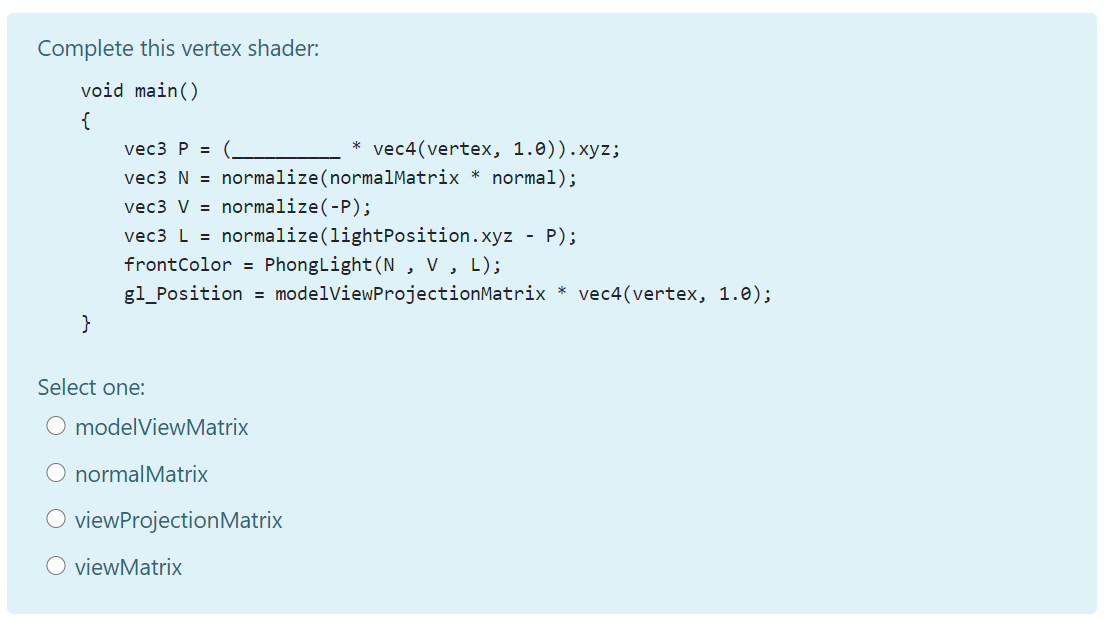} 
\caption{Adding fill-in-the-blanks questions from a list of tokens. One out of the three unique random questions is shown. We put the shader source code inline for a self-contained example, but a better option is to directly include the source code from an existing file.} 
\label{list:complete}
\end{listing*}

Instructors need to check that the selected tokens are not interchangeable, so that randomly chosen alternatives are mutually exclusive. We use extensively this method to ask students to complete code implementing some algorithm studied in class. We often use short token lists, and plausible distractors for these. 

\subsection{Multiple-answer questions}

We do not support this question format since many guides \citep{burton1990} advice against their use because common scoring strategies (all-or-none basis, scoring each alternative independently) have notable disadvantages. 
 
\subsection{Local question preview}
 
The proposed API provides a \minty{preview} method that previews all generated questions in a web browser. Our prototype uses Moodle's CSS templates, so that the preview closely matches that in Moodle (including images, LaTeX formulas\dots). We added  some variations to facilitate preview: question names are shown (question names are hidden to students and not shown in Moodle's previews, but they facilitate search in large question sets); answers to numerical and short-answer questions appear next to the input field; the first choice in multiple-choice questions is the correct one; and drop-down lists in matching questions match subquestion order, so that the correct matching is easy to check.  
Previews are very fast since HTML files are created and browsed locally in the instructor's computer; this contrasts with Moodle's preview, which requires communication with a Moodle server. Previews are based on HTML files that can be easily shared with other instructors e.g. to check, discuss or select questions for an exam. Since HTML files can be searched quickly for arbitrary content (in the question name, text, choices, answers\dots), preview files are also very convenient during online exams to handle student doubts.

 \begin{listing*}[tb]
\begin{minted}[fontsize=\footnotesize]{python}
# Sample script using matplotlib
from quizgen import *
import matplotlib.pyplot as plt
Q = Quiz('listing8.xml')
# Random triangle
A, B, C = [randint(2,9), randint(2,9)], [randint(-9,-2), randint(-9,9)], [randint(2,9), randint(-9,-2)]
T = [A,B,C]
# Some points, in barycentric coordinates
points = [(1/2, 1/2, 0), (1/2, 0, 1/2), (0, 1/2, 1/2), (1/4, 1/4, 1/2), (1/4, 1/2, 1/4), (1/2, 1/4, 1/4),
        (1/5, 4/5, 0), (4/5, 1/5, 0), (0, 1/5, 4/5)]
shuffle(points)
# Pick one random point
bar = points[0]
# Convert to (x,y) coordinates
x, y = bar[0]*A[0] + bar[1]*B[0] + bar[2]*C[0], bar[0]*A[1] + bar[1]*B[1] + bar[2]*C[1]
# Draw plot
plt.clf()
plt.axis('equal')
plt.gca().add_patch(plt.Polygon(T,color='lightblue')) # triangle
for v in T:
    plt.gca().add_patch(plt.Circle(v,radius=0.3,color='blue')) # vertices
plt.text(A[0]+0.5, A[1], s="A") # vertex labels
plt.text(B[0]-0.9, B[1], s="B")
plt.text(C[0]+0.5, C[1], s="C")
plt.gca().add_patch(plt.Circle((x,y),radius=0.3,color='red')) # point
plt.text(x+0.5,y, s="P") # point label
plt.axis('equal')
img=insertPlot(plt)
Q.addMultipleChoice("",f"The barycentric coords of P are:<p>{img}", [bar], points[1:])
\end{minted}
\centering
\includegraphics[width=0.8\textwidth]{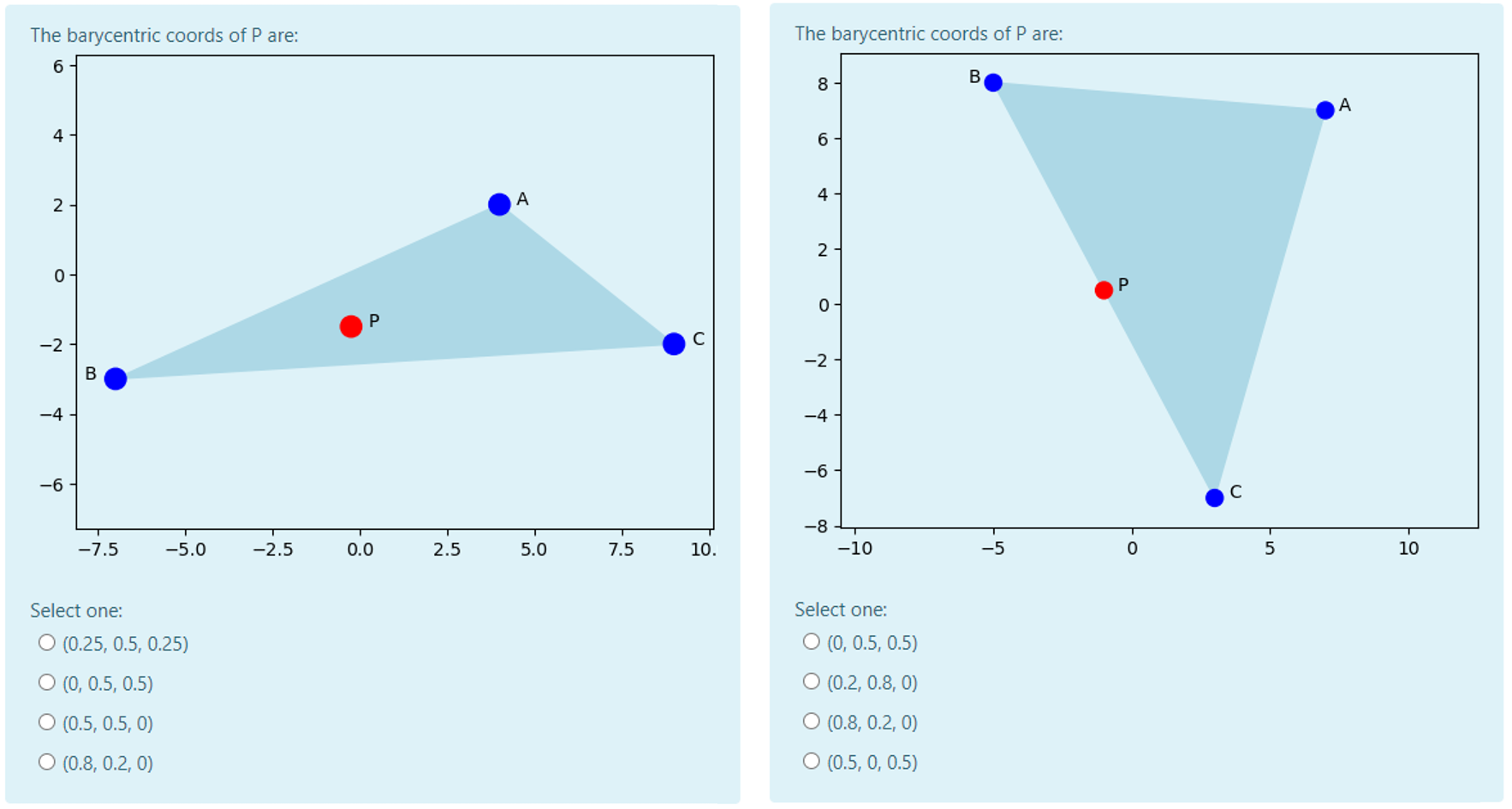} 
\caption{A more complex example using MatPlotLib to draw a random triangle and a random interior point. Drawing the triangles with an external tool and including the images might be faster for this question, but the code that draws the triangle can be reused in related questions with minor modifications.} 
\label{list:bari}
\end{listing*}
 
\subsection{Some useful packages for question generation}

The Python Package Index (PyPI) reports more than 240K Python packages, so a comprehensive review of which packages can be useful for script-based question generation is out of the scope of this paper. Here we just refer to a few Python packages that we found extremely useful for creating questions in our Computer Science courses. 
 
We use the \textbf{random} module to generate random numbers, to take samples from arbitrary collections 
and to shuffle a collection. 
\textbf{SymPy} is very useful for symbolic mathematics (derivatives, integrals, equation solving\dots), and to easily get LaTeX output.
We use \textbf{Matplotlib} to generate plots with random content, \textbf{PIL} for image loading, generation and transformation, and \textbf{pythreejs-libigl}~\cite{libigl,koch2019} for interactive 3D model viewers. Finally, we use \textbf{io.BytesIO} for encoding and embedding dynamic images/videos into HTML. 
 
\section{Results}

In this section we include additional examples to illustrate the possibilities of script-based question generation. Listing~\ref{list:bari} uses MatPlotLib to draw random triangles for questions on barycentric coordinates. Triangle vertices are chosen randomly, as well as the barycentric coordinates of the query point. Listing~\ref{list:fres} also uses MatPlotLib, this time to add questions on the Fresnel equations. Since these questions include dynamically-generated content, reproducing them in current LMS is a much more involved task. 

Listing~\ref{list:det}  illustrates how questions involving matrices can be generated with a small piece of code, thanks to SymPy's features. Since computations (in this case, the matrix determinants) and formatting (LaTeX source for the matrices) are done programmatically, the chance of typing errors in the output questions is highly unlikely. 

\begin{figure}
	\centering
		\includegraphics[width=0.9\columnwidth]{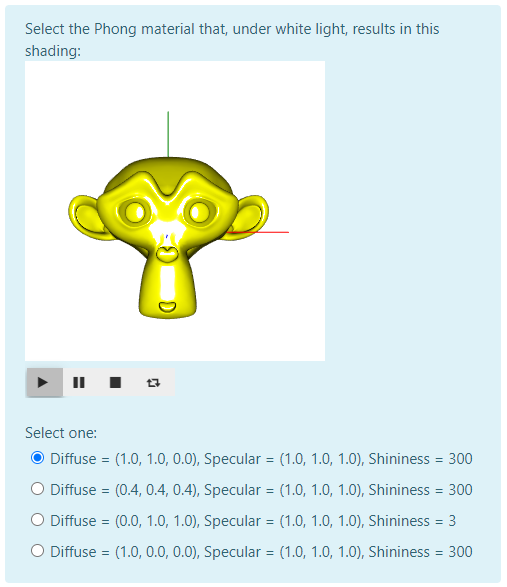}
	\caption{Sample question showing a 3D model with arbitrary material.}
	\label{fig:phong}
\end{figure}

Listing~\ref{list:mesh} illustrates how to create questions on geometric transformations, using an interactive 3D model viewer (supporting zoom, pan and rotation of the model) as part of the question (see accompanying repository for the interactive HTML preview of this question). In the example, rotations and scalings are chosen randomly; notice that these questions might require the students to rotate the model, and thus interaction with the question content is required. The accompanying repository contains further examples: \minty{listing12.py} creates questions on the  Phong reflection model (Figure~\ref{fig:phong}), and \minty{listing13.py} uses Blender's Python API to add questions including renderings with random content.   

\begin{listing*}[tb]
\begin{minted}[fontsize=\footnotesize]{python}
from quizgen import *
import matplotlib.pyplot as plt
import math

# function to plot reflectivity according to Fresnel equations
def fresnelPlot(n1,n2):
    Xs = range(0,91)
    Ys=[]
    for x in Xs:
        R=1
        thetai = math.pi*x/180
        if n1/n2*sin(thetai) < 1:
            thetat = asin(n1/n2*sin(thetai))
            Rs = pow(sin(thetat-thetai)/sin(thetat+thetai),2)
            Rp = pow(tan(thetat-thetai)/tan(thetat+thetai),2)
            R = 0.5*(Rs + Rp)
        Ys.append(R)
    plt.clf()
    plt.plot(Xs, Ys)
    plt.axis([0, 90, 0, 1.05])
    plt.xlabel('Incident angle (degrees)')
    plt.ylabel('Reflectivity')
    plt.grid()
    return insertPlot(plt)
    
# Create quiz
Q = Quiz('listing9.xml')
# Get random values for refractive indices
n1, n2 = 1 + randint(1,15)/10, 1 + randint(1,15)/10
# Add a random question on Fresnel equations
img = fresnelPlot(n1,n2)
choices = [(n1,n2), (n2,n1), (n1+0.5, n2), (n1,n2+0.5), (n1-0.5, n2), (n1, n2-0.5), (n1+0.5,n2+0.5)]
Q.addMultipleChoice("","This plot shows the reflectivity when light hits the interface between media "
+ f"with refractive indices \(\mu_1,\mu_2\):{img} Plausible values for \(\mu_1,\mu_2\) are:", choices)
\end{minted}
\centering
\includegraphics[width=0.99\textwidth]{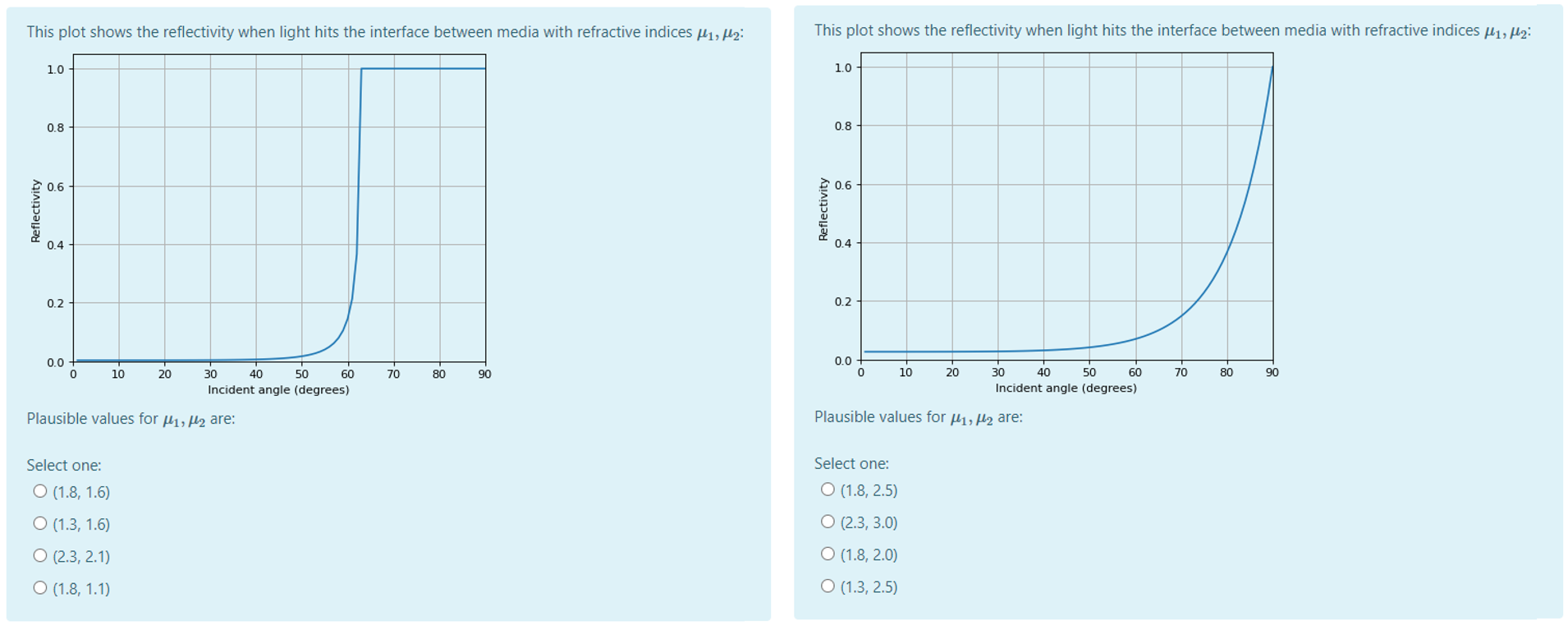} 
\caption{Another example using MatPlotLib. The script generates plots of a function with random parameters.} 
\label{list:fres}
\end{listing*}

\begin{listing*}[tb]
\begin{minted}[fontsize=\footnotesize]{python}
from quizgen import *
from sympy.matrices import randMatrix
Q = Quiz('listing10.xml')
# Get some random 3x3 matrices
matrices = [randMatrix(3,3,0,5) for _ in range(5)]
pairs = [(html(M), M.det()) for M in matrices]
Q.addMultipleChoiceFromPairs("","The determinant of %s is:", pairs)
\end{minted}
\centering
\includegraphics[width=0.8\textwidth]{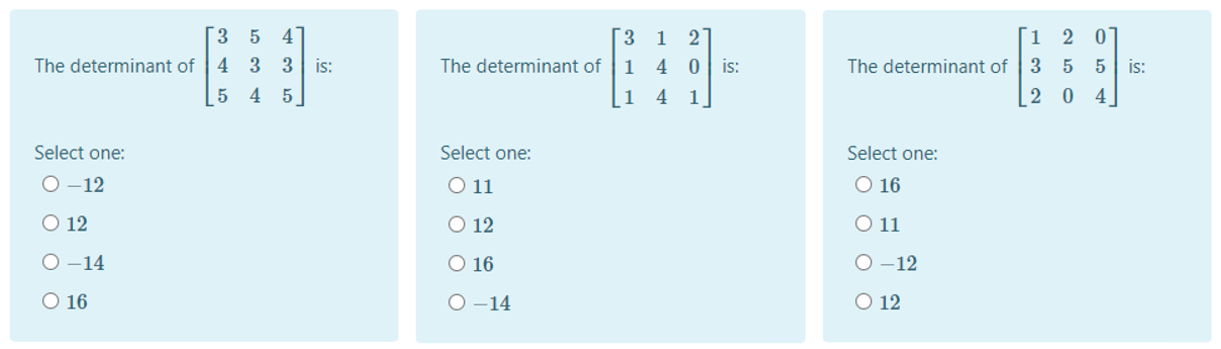} 
\caption{This example creates questions using random matrices.} 
\label{list:det}
\end{listing*}


\begin{listing*}[tb]
\begin{minted}[fontsize=\footnotesize]{python}
# Create a random question on geometric transformations using meshplot
from quizgen import *
from cg_helpers import *

Q = Quiz('listing11.xml')
# Read a 3D model
v, f = igl.read_triangle_mesh("../data/monkey.obj")
# Generate plot
plot1 = plotMesh(v, f).to_html(True, False)
# Generate some rotations
angles = [-45, 45, -90, 90, -135, 135]
rotations = [rotate(axis, a) for a in angles for axis in 'XYZ']
# Generate some scalings
scalings = []
for s in [2, 3, 4]:
    scalings += [scale(1, 1, s), scale(1,s,1), scale(s, 1, 1)]
# Some combinations
transforms = []
transforms += [mult(r, s) for r in rotations for s in scalings] 
transforms += [mult(s, r) for s in scalings  for r in rotations]
# Pick random transforms
shuffle(transforms)
selected = transforms[0]
choices = [item.descr for item in transforms[:4]]
# Transform the model and generate its plot
v = applyTransform(selected, v)
plot2 = plotMesh(v, f).to_html(False,False)
# Add question
question = f"The matrix that transforms the model on the top onto that on the bottom is:{plot1}{plot2}"
Q.addMultipleChoice("", question, choices)
\end{minted}
\centering
\includegraphics[width=0.35\textwidth]{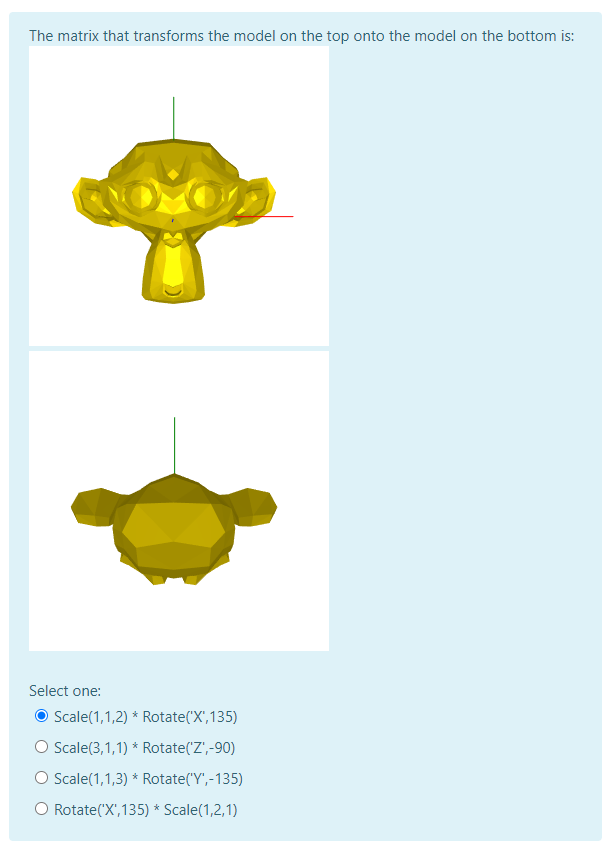} 
\includegraphics[width=0.35\textwidth]{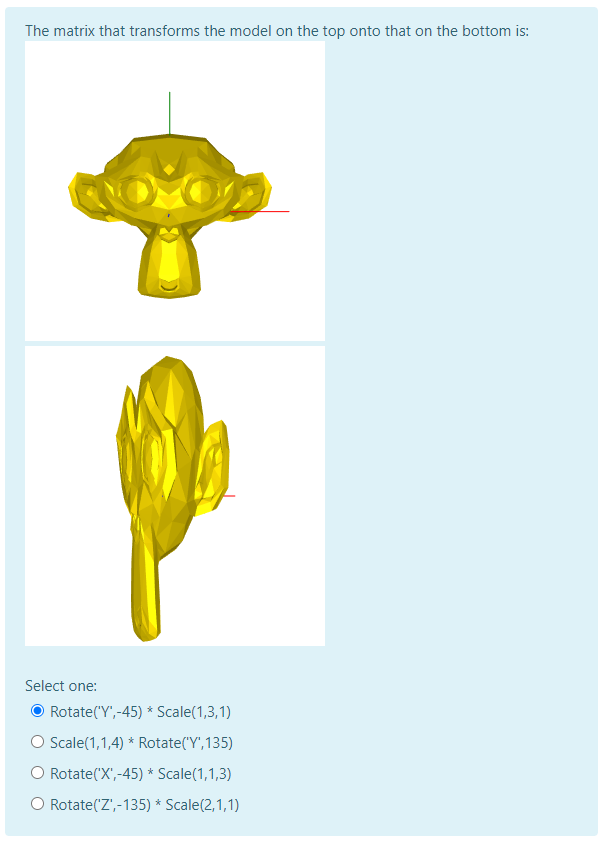} 
\caption{Script that generates a question with an interactive 3D model viewer (see HTML file in accompanying repository).} 
\label{list:mesh}
\end{listing*}

\section{Evaluation}

\subsection{Experiment design}
A formal evaluation of the proposed API through a user study presents serious difficulties. First, we have not released the API yet, so the user base is limited to some Computer Science instructors in our research group. 
Second, question creation performance with GUI-based approaches (e.g. Moodle) is expected to be similar among trained users. That is not the case with a script-based approach, since the prior experience with the API and the programming skills of the instructor might play a crucial role in question entering times, specially when writing scripts that generate programmatically the input for random question generation. 

On the other hand, in a real world scenario, question creation times include the time to design the questions, time to design their script implementation (coding, choosing pairs, choosing distractors\dots) and the time to enter them (typing content in a text editor or in form fields). Since designing the implementation is very sensitive to external factors such as question nature and instructor expertise, the results of such a user study would vary significantly depending on the recruited users and the selected questions. Hence, we have constrained the experiment to the definition of single questions, whereas the entry of families of similar questions, needing little or no extra work when using the proposed API, usually require re-entering each new question when using the GUI-based approach.

We thus decided to evaluate our approach by:
\begin{itemize}
\item Conducting a user-study to measure question \textit{typing} times for \textit{simple} questions (text-based, minimal Python code), using Moodle \textit{vs} a script-based approach. 
\item Fitting a linear model to estimate typing times for simple questions (excluding question \textit{design} and implementation times, as if copying from a paper version of the question). 
\item Reporting measured creation times (script-based) and estimated typing times (Moodle) for a large set of \textit{complex} scripts generating multiple questions. 
\end{itemize}

\subsection{Question creation performance: simple questions}

Here we restrict ourselves to multiple-choice questions, since they are extensively used in quizzes \citep{burton1990} and require varied user input (question, right answer, distractors). 

We asked five trained users (aged 31-55, 3 females, 2 males) to create a quiz by copying a set of 20 text-based questions taken from the examples given by \cite{burton1990}. The total number of characters in the quiz was 5,298; the shortest question included 65 characters whereas the longest one had 651. The average number of characters per question was $n=265$, including stem and the four alternatives. Questions were quite independent from each other, with little options to copy-paste text. 
We asked the participants to configure the questions (both in Moodle 3.8 and with our script-based approach) to common settings: no numbering for the choices, $1/3$ penalty for wrong answers, and default category for all questions. Category creation is much simpler with our approach (using \minty{setCatagory}), whereas Moodle has its own GUI for category creation and management. Considering per-question categories would over penalize Moodle. 

All participants used a commodity PC with a 23'' 16:9 monitor, standard keyboard/mouse, and high-speed (symmetric 100 Mbps) internet connection. Moodle quizzes required communication with a Moodle Cloud server, whereas the script-based approach was based on WinPython 3.8 and the Pyzo IDE. We used an automation tool (AutoHotkey) to record completion times and mouse/keyboard inputs.

\begin{figure}
	\centering
		\includegraphics[width=0.99\columnwidth]{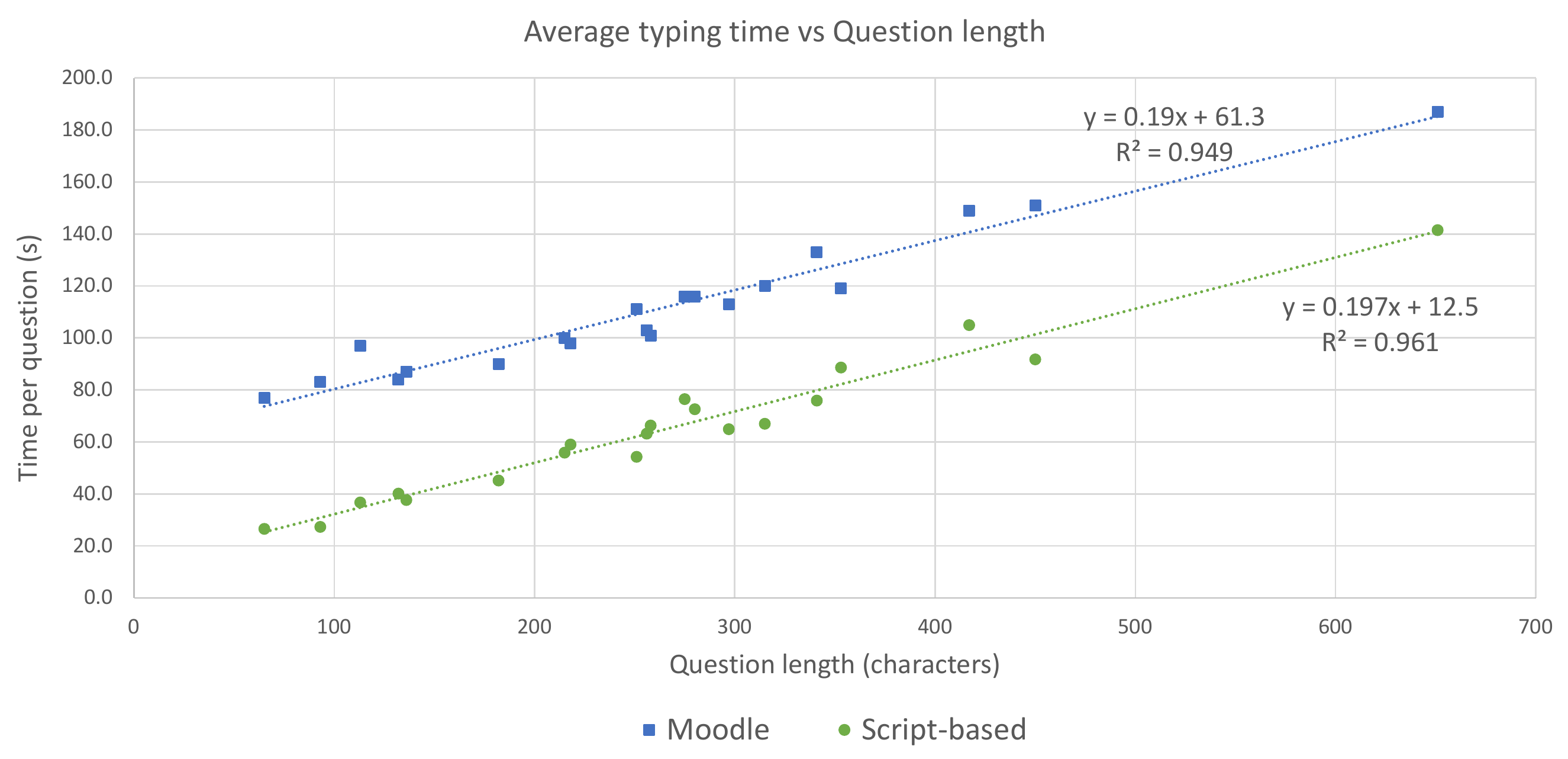}
	\caption{Average times for creating simple questions with Moodle's GUI and our script-based approach.}
	\label{fig:study}
\end{figure}

Figure~\ref{fig:study} shows average completion times for the different questions.  The script-based approach was found to be about 49 seconds consistently faster than Moodle's GUI. 
Low-level input for Moodle (Figure~\ref{fig:moodleform}) depended on user; for example, advancing to the next form field can be achieved either by clicking on the target field, or by pressing the Tab key.  The drop-down lists that Moodle uses to select answer grades (e.g. 100\%, 90\%\dots) also accept different input modalities. In our experiments, users performed on average 23 mouse clicks and 8 scroll operations, per question. This overhead exceeded the extra Python code (about 50 characters) that had to be written for these questions. 
We can conclude that, for typing simple questions, the script-based approach is considerably faster than Moodle, mainly because it adds little overhead when typing the $n$ per-question characters, whereas Moodle involved more mouse/keyboard overhead for moving through the form fields and selecting options from drop-down lists.

We fitted linear models to the collected data to predict question entering time $T$ for these questions (Figure~\ref{fig:study}). The models explained $R^2 > 94\%$ of the variance of completion times, with p-values $< 0.0001.$ 
For questions with $n$ characters, the model estimates $T=0.19n+61.3$\,s for Moodle's GUI, and $T=0.197n+12.5$\,s for the script-based approach, with standard errors $S$=6.2\,s and $S$=5.7\,s, resp. 
Note that these models apply to simple (text-based, minimum Python code) questions, and that we neglect the time needed to design the implementation. 

\subsection{Question creation performance: complex questions}
Question creation involves question design and question typing. But the design of the question itself (the idea of how we may measure the grasp of certain concept or technique by the student) is common to any exam construction technique, in any support, and can therefore be factored out. 
With the script-based approach, instructors must type the question content plus Python code (method names, quotes enclosing text, parenthesis, brackets\dots) and suffer a small overhead in that the \emph{implementation} of the question must also be designed, and perhaps debugged. 
For the simple questions in the previous section, Python code was mostly limited to \minty{addMultipleChoice} calls. When writing scripts that create multiple random questions, script complexity can vary arbitrarily. Some scripts are easy to write (e.g. Listing~\ref{list:low-example}), others require writing LaTeX formulas (e.g. Listing~\ref{list:multi1}), and some others involve significant programming  (e.g. Listing~\ref{list:primes}). We could conduct a user-study to measure times for creating these questions, but creation times would vary significantly depending on users/questions. Furthermore, many of these scripts generate an arbitrary number of random questions, therefore per-question generation times can be as small as we wish by just increasing the number  question variants. 

We thus opted to (a) report observed creation times for the scripts included in this paper, for an experienced Python user; (b) compute a reasonable number of questions to be created by the scripts; when variation is restricted to random values, we set this number to 5 variants; (c) estimate equivalent creation times with Moodle's GUI, using the linear model above.

Table~\ref{tab:results} shows estimated per-question times for our example scripts. This simulation suggests that the script-based approach achieves 2.4-6.3 speed-ups with respect to Moodle's GUI, when creating the random questions that illustrate this paper. 
These times are meant to be just a rough estimate of what instructors with Python experience might expect from using our approach \textit{vs} Moodle's GUI. Notice that actual creation times will increase significantly if the instructor has no experience with Python or specific packages. Conversely, creation times can also be reduced by re-using and factorizing code. For example, Listing ~\ref{list:det} on determinants can be trivially modified to ask for the matrix inverse or its SVD. This possibility of re-using code and creating \textit{macros} extremely simplifies the creation of large question banks. 

\begin{table*}
\footnotesize
\centering
\begin{tabular}{@{}rrrrrrr@{}}
\toprule
\multicolumn{3}{c}{Script-based}                                                             & \multicolumn{4}{c}{Moodle GUI}                                                                                              \\ \midrule
\multicolumn{1}{l}{Listing} & \multicolumn{1}{l}{Time (s)} & \multicolumn{1}{l}{\#Questions} & \multicolumn{1}{l}{Avg length  (chars)} & \multicolumn{1}{l}{Estimated question time (s)} & \multicolumn{1}{l}{Time (s)} & \multicolumn{1}{l}{Speed Up}\\ \midrule
1   & 160 & 6 & 180  & 95  & 570   & 3.6                        \\ \midrule
3   & 198 & 11 & 70  & 74  & 814  & 4.1                         \\ \midrule
4   & 220 & 15 & 170 & 93  & 1395  & 6.3                         \\ \midrule
5   & 309 & 7 & 400  & 137 & 959   & 3.1                    \\ \midrule
6   & 390 & 5 & 85   & N/A & N/A   & N/A                 \\ \midrule
7   & 182 & 3 & 450  & 146 & 438   & 2.4                         \\ \midrule
8   & 440 & 5 & N/A  & N/A & N/A   & N/A                       \\ \midrule
9   & 512 & 5 & N/A & N/A & N/A   & N/A                       \\ \midrule
10   & 77 & 5 & 40   & 68  & 340   & 4.4                        \\ \midrule
11   & 480 & 5 & N/A   & N/A  & N/A   & N/A                        \\ \bottomrule
\end{tabular}
\caption{Comparison of creation times for script-based \textit{vs} Moodle's GUI. 
For each listing in the paper, we provide the measured creation times for an instructor with Python experience (times include question design, coding and typing), and a target number of questions. We also estimate the time to create these questions with Moodle's GUI by computing average per-question lengths (except for those containing images), estimated per-question times as predicted by the lineal models above, and total times for the target number of questions. The last column reports the speed-up factor, calculated as the estimated Moodle's GUI time divided by script-based time. 
We did not estimate Moodle times for questions with non-trivial distractors (Listing 6) or involving images with random content (Listings 8, 9, 11). }
\label{tab:results}
\end{table*}

\subsection{Question maintenance performance}
A further advantage of a script-based approach is the automation of maintenance operations. Some large-scale question modifications take considerable amounts of time in GUI-based LMS. For example, language translation might involve duplicating all questions and going through all of them, field-by-field, to translate them. Another example is changing wrong-answer penalties for multiple choice questions. For example, one might decide not to penalize wrong answers (or, conversely, penalizing them e.g. by the typical 1/(n-1) penalty. In current LMS, this means going through all these questions, locating the associated form fields, and changing the penalties of the wrong answers. For large question banks, this tedious task might take hours and is prone to errors.

The following modifications can be completed by the instructor in constant time (less than one minute) using the proposed API, whereas for Moodle they require going through all the questions to edit them (linear time): 

\begin{itemize}
    \item Replace some text in all questions (e.g. 'Flux' by 'Radiant flux').
    \item Modify wrong-answer penalties for all questions (e.g. -33.3\% to 0\%).
\end{itemize}

\section{Conclusions and future work}

The major motivation of a script-based approach is the ability to create large, rich question banks efficiently. STEM instructors using quizzes might adopt our approach at different levels. The simplest one involves using the API with inputs (answer lists, distractor lists\dots) written manually in the Python code. This approach already saves a significant amount of time (about 50\,s per question), and requires minimal programming skills (knowing how to enclose Python strings and lists, and calling a few API methods). We believe this should be accessible to most STEM instructors. 
A second level involves creating questions with random parameters. The amount of Python code needed to generate random parameters is small, but often some extra code is needed to compute the answers, to choose proper values for parameters and answers, and to create plausible distractors. Still, many STEM instructors should be able to test the API at this level with little effort.
The third level consists in generating complex input programmatically. Some Python packages greatly simplify this task (e.g. SymPy in  Listing~\ref{list:det}), but depending on the questions to be generated, instructors would require additional programming skills and, in some cases, learning how to use new Python packages. We guess that some instructors with experience e.g. in Python/MatLab might want to try this approach, specially if they need to create/redefine large question banks when moving from traditional to online quizzes.

Major benefits of our approach are (a) faster creation of questions in a variety of fields; (b) easy creation of dynamically generated content (images, plots, interactive 3D model viewers...). Additional benefits include (c) simpler maintenance operations, and (d) possibility of collaboration and version control (e.g. git). 
Unlike some proprietary SDKs which focus exclusively on education, instructors adopting our approach might acquire, as a side benefit,  additional programming skills with one of the major languages in scientific computing~\cite{virtanen2020scipy}.

Our approach has some limitations though. Instructors need to feel comfortable with Python coding. We neglected script debugging times, although this was not an issue in our experience (after writing scripts generating thousands of questions), thanks to the immediate feedback provided by the Python IDE and the preview option of the API. Although many STEM instructors are familiarized with LaTeX and basic HTML, they might prefer WYSIWYG editors such as those integrated in LMS. The fast preview option in our implementation, and Python packages such as SymPy, partially mitigate this limitation. Another limitation is that question variants are compactly represented in the Python scripts, but have duplicated content in the exported XML question file and thus in LMS question bank. 
Our approach does not compete, but complements, other question types (e.g. WIRIS quizzes). Actually, since WIRIS questions are also imported/exported as Moodle's XML format, we are working on supporting the generation of questions that benefit from runtime expression equivalence. Finally, if the proposed API gets acceptance among users from LMS communities, we plan to conduct a TAM (technology acceptance model) study on script-based quiz generation.  
\paragraph*{Source code git repository} \url{https://gitrepos.virvig.eu/docencia/QuizGen/}

\section*{Acknowledgments}
The author would like to thank Dr. Alvar Vinacua for his assistance with the repository setup and his valuable and constructive suggestions on the manuscript. 
This work has been partially funded by the Spanish Ministry of Economy and Competitiveness and FEDER Grant TIN2017-88515-C2-1-R.

\bibliographystyle{unsrt}
\bibliography{refs}

\end{document}